\documentclass[journal]{IEEEtran}

\ifCLASSINFOpdf
  \usepackage[pdftex]{graphicx}
\else
\fi

\usepackage{amsmath,amssymb}
\usepackage{amsfonts} % is this going to be a problem..?
\usepackage[justification=justified,font={normalfont,singlespacing,footnotesize},singlelinecheck=off]{subcaption}

\usepackage{xcolor}
\usepackage[normalem]{ulem}
\usepackage{xifthen}
\usepackage[textsize=small]{todonotes}

\usepackage{color,soul}

\usepackage[english]{babel}
\usepackage{blindtext}

\newcommand{\TODO}[1]{\textsf{\textbf{[TODO: }\textsl{#1}\textbf{]}}}

\newcounter{DefCounter}

\newcounter{LemCounter}

\newcounter{ThmCounter}
\usepackage[justification=centering]{caption}

\usepackage[hyphens]{url}
\usepackage[breaklinks=true]{hyperref}

\begin{document}

\title{Engineering Economics in the Conflux Network}

\author{\IEEEauthorblockN{Yuxi Cai\IEEEauthorrefmark{1}, Fan Long\IEEEauthorrefmark{2}, Andreas Park\IEEEauthorrefmark{3}, Andreas Veneris\IEEEauthorrefmark{1}\IEEEauthorrefmark{2}}

\IEEEauthorblockA{\IEEEauthorrefmark{1}Dept. of Electrical and Computer Engineering, University of Toronto }

\IEEEauthorblockA{\IEEEauthorrefmark{2}Dept. of Computer Science, University of Toronto}

\IEEEauthorblockA{\IEEEauthorrefmark{3}The Rotman School of Management, University of Toronto }
 
\texttt{caiyuxi@ece.utoronto.ca, fanl@cs.toronto.edu}

\texttt{andreas.park@rotman.utoronto.ca, veneris@eecg.toronto.edu}\\
}

% Copyright notice for camera-ready copy
\IEEEoverridecommandlockouts
%\IEEEpubid{\makebox[\columnwidth]{978-1-5386-5541-2/18/\$31.00~\copyright2019 IEEE \hfill} \hspace{\columnsep}\makebox[\columnwidth]{ }}

\markboth{
%Journal of XYZ,~Vol.~14, No.~8, May~2019
}%
{Author1 \MakeLowercase{\textit{et al.}}: Title}



\maketitle

% Render copyright notice for camera-ready
 \IEEEpubidadjcol

\begin{abstract}

% Q. Why was this work undertaken? What is the motivation?
% A. 

% Q. What did we do, and how?
% A. 

% Q. What did we find?
% A. 

% Q. How well does it work?
% A. 

%This is placeholder text does not come anywhere close to capture the sheer awesomeness of our work. \blindtext[1]
Proof-of-work blockchains need to be carefully designed so as to create the proper incentives for miners to faithfully maintain the network in a sustainable way. 
This paper describes how the economic engineering of the Conflux Network, a high throughput proof-of-work blockchain, leads to sound economic incentives that support desirable and sustainable mining behavior.
In detail, this paper parameterizes the level of income, and thus network security, that Conflux  can generate, and it describes how
this depends on user behavior and ``policy variables'' such as block and interest inflation. 
It also discusses how the underlying economic engineering design makes the Conflux Network resilient against double spending and selfish mining attacks.
\end{abstract}


\IEEEpeerreviewmaketitle

%auto-ignore

\section{Introduction}
%Decentralized oracle: needs and benefits

%usage of blockchain as crytocurrency and dapp platform
Blockchain technology allows peer-to-peer electronic value transfers without the involvement of trusted third parties.
Trust about the completion of financial transactions in the traditional world of finance rests on the economic principle that the trusted (third) party has too  much to lose from negligence or cheating (e.g., regulation penalties, loss of reputation,
reduced future income revenue streams, etc). 
Blockchain networks like Bitcoin~\cite{nakamoto2012bitcoin} and Ethereum~\cite{buterin2014ethereum} use a different mechanism that decentralizes the financial ledger among all participants of the network. As long as a majority of the network participants behave honestly, the fidelity of the ledger is guaranteed by their underlying consensus algorithms.

Both Bitcoin and Ethereum employ Proof-of-Work (PoW) schemes to secure the networks and to defend against Sybil attacks~\cite{sybilAttack}. 
In PoW, miners compete to solve a cryptographic puzzle that requires excessive computational random guessing (aka ``work''). 
The winner has the right to generate a new block and receives a reward for generating the block in the native crypto-currency. 
The PoW mechanism accomplishes several things simultaneously: it creates consensus as to who proposes a new block, and it 
introduces sufficient uncertainty as to who gets to propose one next. 
This implicit randomness is subject to resource expenditures,
{\em i.e.,} the more one spends/works, the more likely that person 
wins. 
Since PoW involves expenditure of resources, the rewards that miners receive are directly related to the security of the network: the more miners earn, the more they computationally compete to secure the network.

In PoW, participants in the network agree on the ``longest chain'' as the transaction history of the blockchain ledger.
To make the transaction history secure and irreversible, new blocks are expected to be appended at the end of the longest chain to make it even  harder and hence economically costly to revert~\cite{nakamoto2012bitcoin}. Notably, users
have to wait for a sufficient number of blocks after the transaction so that the state change is irreversible.  
This serial processing of blocks puts severe constraints on network throughput, which limits the usability of the platforms in day-to-day real-life monetary transactions. 

Aside from the performance challenge of the limited throughput when compared to traditional financial networks
(VISA, SWIFT, etc), blockchain networks face two economic challenges. 
First, the long-term economic sustainability of blockchain networks like Bitcoin and Ethereum remains unclear. Bitcoin is currently secure because miners receive substantial block rewards. 
Several  studies argue, however, that as the block rewards phase out, Bitcoin will be much more vulnerable to double spending attacks~\cite{absentBlockReward}. 
Second, the cost of maintaining a blockchain network grows as the network  adds users and transactions. 
For example, Ethereum supports the deployment of decentralized applications through the execution of ``smart contracts.'' 
Users pay only a one-time inclusion payment for their smart contract code, but following this, the smart
contract occupies state storage without further costs. As such, inactive smart contracts (that is, the majority of
smart contracts in Ethereum today) lead to inefficient usage of space and drive up the cost of
maintaining the network.

%Another notable vulnerability of those existing blockchain networks with a sequential ledger is the easiness of a fiarness atacks. 
%\TODO{I dislike this paragraph and I don't understand the point as it relates to this paper. It's not very precise.}
Notably, blockchain networks with sequential ledgers are also vulnerable to fairness attacks. A network participant with more than 23.21\% computation power can employ a special block mining strategy to launch selfish mining attacks to obtain block rewards that are disproportional to its computation power~\cite{sapirshtein2015optimal}. 
Because PoW mining is a winner-take-all game for miners to compete on including blocks into the longest chain,
a malicious participant can strategically withhold some of her mined blocks to gain the advantage of exclusively mining on the longest chain~\cite{eyal2013majority}. 
Such fairness attack strategies put small miners into a disadvantage and may cause the blockchain network to become increasingly centralized, therefore exploiting fairness and undermining the fidelity of the blockchain ledger.%\TODO{so what?} 

Conflux~\cite{li2018scaling}  is a new PoW network  with a Turing-complete smart contract language similar
to this of Ethereum. The Conflux network provides significant performance improvements with its processing 
of parallel blocks in a directed acrylic graph (DAG) structure, which lowers
confirmation times and increases transaction throughput substantially. 

%forcing the figure to show on the top of page 2
\begin{figure*}[t]
\centering
\includegraphics[width=15cm]{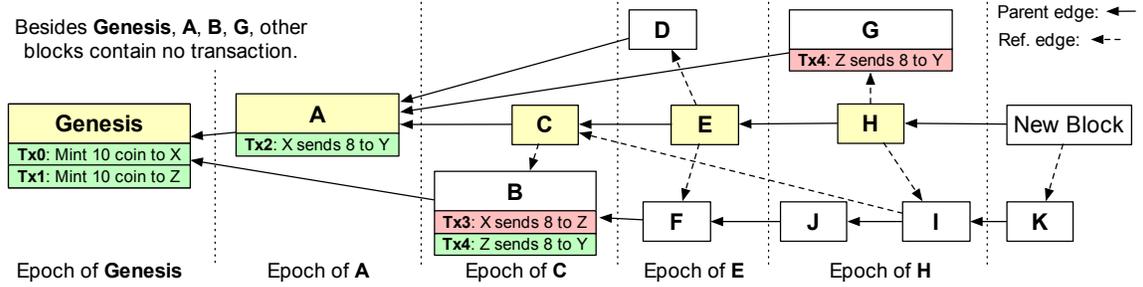}
% where an .eps filename suffix will be assumed under latex, 
% and a .pdf suffix will be assumed for pdflatex; or what has been declared
% via \DeclareGraphicsExtensions.
\caption{TreeGraph structure example in Conflux. Yellow blocks corresponds to the pivot chain.}
\label{treegraph}
\end{figure*}

This paper focuses on the economic engineering and the incentive mechanism design of the Conflux network. 
To address the space congestion  challenge, Conflux requires users to stake native tokens as storage bonds to occupy space, which   implicitly creates a disincentive to occupy space unnecessarily. 
The disincentive stems from the payment of interest on existing tokens in the system. 
The interest on the storage bonds is payed to miners instead of the users to create a long term income to the miners. 
To address the fairness attack challenge, Conflux assigns the block reward in a way that eliminates the winner-take-all characteristic of mining.  
Instead of competing for the longest chain, miners in Conflux receive block rewards for all the blocks that they generate, albeit  with some penalty mechanisms that encourage following the consensus protocol. 
Competing blocks are jointly penalized so that selfish mining is not profitable and different miners are incentivized to cooperate along the protocol to keep the network stable and secure.

This paper makes the following contributions: 1) we analyze the economic impact of the proposed token rules for Conflux; 2) we show that an optimal selfish mining strategy is not profitable on Conflux; 3) we show that a double-spending attack on Conflux is more difficult compared to legacy blockchain networks with sequential ledgers. 

The remainder of the paper is organized as follows. 
Section~\ref{sec:prelim} presents Conflux with the focus on its economic and incentive mechanisms.
Section~\ref{sec:calib} derives a calibrated economic model for miner income to analyze the long term sustainability of Conflux. 
Section~\ref{sec:limits} shows that Conflux has a stricter requirement for potential attacker than sequential systems. 
Section~\ref{sec:conclusion} concludes.

\section{An Overview of the Conflux Network}\label{sec:prelim}

The section presents an overview of the Conflux network~\cite{li2018scaling, confluxSpec:2020}. 
Similarly to Ethereum,  Conflux operates with an account-based model that every normal account associates with a balance and each smart contract account contains the corresponding byte codes as well as an internal state.
Conflux supports a modified version of Solidity (the main contract language in Ethereum) and Ethereum Virtual Machine (EVM) for its smart contracts, so that smart contracts from Ethereum can migrate to Conflux easily.

A transaction in Conflux refers to a message that initiates a payment transaction, or deploys/executes smart contract code. 
Each block consists of a list of transactions that are verified by the proposing miner. 
Each node maintains a pool of verified, received transactions that have not yet been included in a block. 
Miners compete with one another by solving PoW puzzles to include transactions into blocks. 
Similar to Bitcoin and Ethereum, Conflux adjusts the PoW difficulty so as to maintain a stable block generation rate. 
Each node also maintains a local state constructed from the received blocks. 

\subsection{Consensus with TreeGraph}

The Conflux consensus algorithm operates with a special directed acyclic graph (DAG) structure called TreeGraph.
Figure~\ref{treegraph} presents an example of the TreeGraph structure that the Conflux consensus algorithm uses to organize blocks. 
Unlike Ethereum which only accepts transactions on a single chain into its ledger, the Conflux consensus algorithm safely incorporates and processes transactions in all concurrent blocks~\cite{li2018scaling,confluxSpec:2020}.
There are two kinds of edges between blocks, {\em parent} edges and {\em reference} edges. 
Each block (except the genesis) in the TreeGraph has exactly one parent edge to its chosen parent block ({\em i.e.,} solid edges in Figure~\ref{treegraph}). 
Each block can also have multiple reference edges to refer previous blocks ({\em i.e.,} dotted edges in Figure~\ref{treegraph}). 
All parent edges form a tree embedded inside a directed acyclic graph (DAG) of all edges. 

At a high level, Conflux uses the novel Greedy Heaviest Adaptive SubTree (GHAST)~\cite{confluxSpec:2020} algorithm, which assigns a weight to each block according to the topologies in the TreeGraph. 
Under this weight assignment, there is a deterministically heaviest chain within the graph called {\em pivot chain}, which corresponds to the relatively most stable chain from the genesis to the tip of the parental tree. 
For example, in Figure~\ref{treegraph} the pivot chain contains blocks Genesis, A, C, E, and H. 
To generate a new block, a miner will choose the last block of the pivot chain as the parent of the new block. 
The new block will also reference all blocks that have no incoming edge (parent or reference edges) as shown in Figure~\ref{treegraph}. 
This is similar to the idea of extending the longest chain. 
The goal is to make the pivot chain even more stable so that everyone in the network can converge and agree on the same pivot chain.

Parent edges, reference edges, and the pivot chain together enable Conflux to split all the DAG blocks into {\em epochs}. 
As shown in Figure~\ref{treegraph}, every block in the pivot chain corresponds to one epoch. 
Each epoch contains all blocks that are reachable from the corresponding block in the pivot chain via the combination of parent edges and reference edges and that are not included in previous epochs.
Conflux then derives a deterministic total order of blocks as follows: 1) first sort blocks based on epochs (e.g., A is ahead of F); 2) for blocks in the same epoch, sort them based on the topological order (e.g., J is ahead of H); 3) use block id to break ties. 
Because all participants will converge and agree on the same pivot chain over time, they will also derive and agree on the same total order of blocks. 
Participants therefore process all transactions based on the derived block total order. 
For duplicate and conflicting transactions, Conflux will only process the first occurrence and discard the remaining as no-ops.

Experimental results have shown that Conflux is capable of processing $3,200$ tps for simple payment transactions~\cite{li2018scaling}, at least two orders of magnitude higher throughput than Ethereum and Bitcoin. 
The improvement in throughput is a result of the DAG structure and the consensus algorithm, so that the network can operate with a much faster block generation rate, no forks are discarded, and with a higher utilization of block space. 
According to the technical specification~\cite{confluxSpec:2020}, the main net of Conflux (expected in the second quarter
of 2020) will run under a fixed block generation rate at two blocks per second. 
The daily block generation rate is therefore $60\cdot60\cdot24\times 2=172,800$ blocks per day.

\subsection{Conflux Token and Interest}
\label{sec:token}

There is a unique native token on the Conflux network, hereafter referred to as {\em CFX}.
Each CFX contains $10^{18}$ \textit{Drip}, the minimum unit of the native token.
CFX plays a similar role as the native tokens in the Ethereum networks. 
Users submit a contract with a gas limit and a gas price where the latter is denominated in~CFX.

The issued CFXs exist in two forms: {\em liquid} and {\em illiquid}. 
In the liquid form, they can be immediately transferred/used on the Conflux network while the user does not receive any interest payment. 
Illiquid tokens cannot be transferred unless they are ``unlocked''. 
There are two forms of illiquid tokens: 
\begin{enumerate}
    \item \textbf{Locked tokens:} Tokens can be locked up so as to earn the user interest, and 
    \item \textbf{Storage bonds:} Tokens can be set as storage bonds to rent space on the network ({\em e.g.,} for running smart contracts). The required storage bonds are proportional to the amount of space that the contract occupies.
\end{enumerate}

All illiquid CFXs generate interest in the Conflux network. 
Users receive tokens from locked tokens.
Miners receive the interest payment from storage bonds as maintenance fees for storing contract data. 

In this paper, we use $r_c$ for the system base interest rate, expressed in annual terms, and interest is compounded per block. 
Therefore, a user that stakes for $b$ blocks receives an interest payment of
\[\left(1+\frac{r_c}{63,072,000}\right)^b-1\] 
per staked token.
For instance, if the annual interest is $r_c=2\%$, a user that stakes for 15,768,000 blocks (around a fiscal quarter) will receive interest of around $.5\%$ per staked token. 
In  calculations,   interest payments are rounded \textit{down} to the nearest one ($1$) drip. 

\iffalse
For a nominal rate of $r_c\%$, the  annual rate is:
\[\text{effective annual rate}=\left(1+\frac{r_c}{63,072,000}\right)^{63,072,000}-1.\]
For instance, for $r_c=2\%$, the  $\text{effective annual rate}\approx 2.04\%$.
\fi

The economic mechanism is straightforward: paying interest leads to an increase in the number of tokens (the ``monetary base'').
Since users only receive payments from illiquid tokens, the interest payments implicitly shift value from those who do not stake to those who stake.

\subsection{Mining Rewards}
Network maintainers (miners) of the Conflux network receive income from three sources: transaction fees, block rewards, and interest income that arises from users ``renting'' space on the blockchain, as follows. 
 
\begin{enumerate}
    \item \textbf{Transaction Fees:} In the long run, transaction fees will make up the major portion of rewards because of the higher transaction throughput of the network. With many transactions, even very small fees add up to a substantial income. 
    
    \item \textbf{Block Rewards:} As the common practice in PoW networks, the miner of a block receives a coin-base reward. 
    Over time, these rewards increase the monetary base and lead to inflation. 
    %For the rest of the paper, we denote block reward with $B$ in the unit of CFX per block. We denote this block-reward driven inflation rate by $r_b$. 
    Ignoring any market-driven price changes,  economically coin-based rewards are a transfer of wealth from existing CFX holders collectively to the winning miner.
     
    \item \textbf{Storage interest:} As mentioned in Section~\ref{sec:token}, when tokens are used as bonds for storage, the interest paid on those tokens is passed on to miners.
    Similar to the block reward, the total amount of interest from storage bonded tokens will be distributed according to the block weights for each miner. 
\end{enumerate}

\subsection{Anti-cone Penalty Ratio}

The final mining reward of a block is modified by an anti-cone penalty ratio in Conflux. 
Suppose the base reward of a block $b$ combining transaction fees, block reward, and storage interest payment is $B_0$. 
In this paper, we define the final reward of $b$ as:

\begin{displaymath}
B_0 \cdot \max\left\{0, 1 - \left(\frac{|\mathrm{Anticone}(b)|}{100}\right)^2 \right\}
\end{displaymath}
In the above, $\mathrm{Anticone}(b)$ denotes the set of blocks that are not in the past sub-graph of $b$ ({\em i.e.,} 
reachable via parent and/or reference edges from $b$) nor in the future sub-graph of $b$ ({\em i.e.,} 
reachable via parent and/or reference edges to $b$). 
For example, $\mathrm{Anticone}(F) = \{D, G\}$ in Figure~\ref{treegraph}. 
Because the anti-cone  of a block may keep growing, $\mathrm{Anticone}(b)$ here only includes blocks that are within 10 epochs after the epoch where $b$ resides in. 
Note that for simplicity, we exclude difficulty adjustment from the consideration of the formula and assume the difficulty remains constant. We refer the interested reader to~\cite{confluxSpec:2020} for a comprehensive description of
difficulty adjustment. 

For a new block, the base reward is the maximum block reward the generator can possibly receive. 
For every anti-cone block of the new block, a portion of the block reward will be deducted till zero. 
Intuitively, this block reward formula encourages the generator to conform with the honest behavior as defined by the consensus protocol. 
It encourages the generator to refer as many blocks as possible to avoid unreferenced anti-cone blocks. 
It also encourages the generator to propagate the new block as soon as possible to avoid anti-cone blocks due to network delay. 
Unlike the winner-take-all mining game  for the longest chain in Bitcoin, all blocks in Conflux  receive block rewards and miners who  cooperate  with one another  minimize the anti-cone. 
This makes Conflux secure against selfish mining attacks which exploit the winner-take-all nature of Bitcoin mining~\cite{eyal2013majority}.

\begin{table}[ht]
\caption{List of Symbols\label{notation}}

    \centering

\begin{tabular}{ll}
\hline\\
Symbol & Meaning \\\\
\hline\\
$G$ & genesis tokens\\
$D$ & number of seconds in a day, $60\times60 \times 24$\\
$d$ & days since main-net launch \\
$B$ & block reward \\
$b(d)$ & block rewards per day \\
$r_b$ & annual inflation rate from block rewards \\
$u(d)$ & user uptake rate $\in(0,1)$) \\
$u^\text{ETH}$ & estimated user uptake rate based on Ethereum\\
$u^{\text{fast}}(d),~ u^{\text{slow}}(d)$ &  $u^\text{ETH}$ advanced/delayed by 180 days respectively \\
$T(d)$ & number of transactions on day $d$ \\
$f$ & average transaction fee \\
$F(d)$ & total transaction fees paid to miners on day $d$\\
$\alpha$ & fraction of tokens that are locked \\
$r_c$ & annual rate of inflation due to interest payments\\
$R$ & daily interest rate for compound transactions \\
$\gamma(d)$ & fraction of gas used by computations \\
$\beta$ & required fraction of tokens as storage bonds \\
$I(d)$ & interest income from storage bonds for miners \\
$p(d)$ & inflation adjusted price on day $d$ \\
$G(d)$ & number of coins outstanding on day $d$\\
$m(d)$ & total revenue for miners on day $d$ \\
$\bar{m}(d)$& total miner revenue averaged over 1 year \\\\\hline

\end{tabular}

\end{table}

%auto-ignore

\section{Calibrated Economic Models}\label{sec:calib}
Miners are essential to the security of the network, and the computing power they contribute to secure the network is (empirically) increasing in the revenue that they can earn.
In this section, we develop an economic
approach to determine the expected revenues that miners gain from participating in Conflux. 
We calibrate this model based on our technical specification as well as historical data
from Ethereum given the similarity in network features. 
As a reference, Table~\ref{notation} outlines the symbols used in this section. 
Each of the following
first three subsection 
discusses one component of the miner reward and the last subsection presents the overall expectation. 

\subsection{Block Rewards}
Assuming a  constant mining rate of~$2$~blocks per second~\cite{confluxSpec:2020},  there are $2D\cdot 365$ blocks mined per year by Conflux. 
As such, if $B$ denotes the number of newly minted tokens created as block reward to the miners, the  system needs to issue  $B\cdot 2D \cdot 365$ new tokens annually as block rewards.
 Blocks rewards   increase the monetary base and create inflation. 
 Specifically, Conflux's objective is to set  the block reward based on  an annual  inflation target rate of $r_b\in(0,1)$. 
Therefore, for target value $r_b$, the block reward must solve
\[B\cdot 2D \cdot 365\equiv G\cdot r_b ~~\Leftrightarrow~~B=\frac{Gr_b}{730 D}.\] 
Overall, on any given day $d$, total block reward $b(d)$ is:
\begin{equation}\label{blockrewards}
    b(d)=Gr_b/365. %\frac{Gr_b}{365}.
\end{equation}

\subsection{User Uptake and Transaction  Fees}\label{uptake}
To model the expected transaction fees, we first develop a model for the user uptake rate modeling after Ethereum. Network user adoption directly relates to the demand for transactions and smart contract computation, the fees paid by users, and the storage interest distributed to miners. 
%We discuss the latter quantities in later subsections as here our objective is to develop a user adoption model.

\begin{figure}[t]
\begin{center}
\includegraphics[width=7cm]{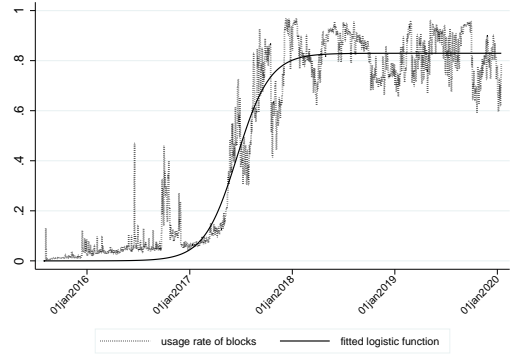}
\caption{Adoption of Ethereum: a fitted logistic function\label{eth_uptake_fig}}
\end{center}
\end{figure}

A common feature of new technologies is that  their adoption follows a S-shaped pattern with slowly increasing usage early on and then a sudden jump of activity~\cite{innovation}. The user adoption rate in Ethereum, as depicted in Figure \ref{eth_uptake_fig}, indeed shows such a feature. We plot the average fraction of space (or gas) occupied in a block as a function of time, based on data from \cite{etherscan}.

%Conflux and Ethereum have many similar features and we will therefore use Ethereum's historical adoption data  to estimate what we believe to be a reasonable uptake rate for Conflux.

We characterize the Ethereum's user uptake sigmoid curve with a logistic function which has the form: 
\begin{equation}\label{logistic}
    %Y = \frac{b_0}{  1 + b_1 \cdot e^{-b_2 \cdot X}},
    y = \frac{\xi_0}{  1 +  e^{-\xi_1 \cdot (x-x_0)}},
    \end{equation}
In the equation above,  $y$ is the uptake rate at time $x$, $\xi_0$ is the maximum value for uptake,  $\xi_1$ is the growth rate, and $x_0$ is the time-value of when the curve reaches 50\% of its maximum value (formally, the value of the horizontal axis at the sigmoid's midpoint).
%Our estimation results are in Table \ref{eth_est}, Figure \ref{eth_uptake_fig} also plots the fitted function.
Following the results for the non-linear least squared regression of~(\ref{logistic}) we obtain a user uptake rate function $u^{\text{ETH}}(d)$ as follows: 
\begin{equation}\label{eth_cali}
    u^{\text{ETH}}(d)=\frac{0.83}{1+ e^{-0.017\cdot (d-690)}}.
\end{equation}

\iffalse
\begin{table}[ht]
\caption{Logistic Curve-Fitting for Ethereum's Uptake Rate}\label{eth_est}

    \centering
    \begin{tabular}{llll}
    \hline\\
 & $\xi_0$ & $\xi_1$ & $X_0$ \\
 \\\hline\\
Estimate & 0.83***\footnotemark & 0.02*** & 690.54*** \\
 & (244.80) & (30.81) & (316.07) \\
 \\
Observations & 1,631 & 1,631 & 1,631 \\
R-squared & 0.978 & 0.978 & 0.978 \\
\\\hline
\end{tabular}
\end{table}

\footnotetext{For all tables, *, **, *** indicate significance at the 10\%,  5\%, and 1\% levels. T-stats are in parentheses.}
\fi

It is notable that blocks can theoretically be filled up to 100\% of the gas limit, yet the estimate for $\xi_0$ indicates that the Ethereum blockchain's  usage rate  currently maxes out at  83\%.
There could be three explanations for this. 
First,  miners may collude so to not include transactions that offer low transaction fees. 
Next,  the 83\% usage rate is the  ``technological'' upper bound of what miners can actually include  accounting for  validation and transaction submission latency.
Third, it is  possible  that once the network becomes congested, users no longer  send new transactions to the network because of the long delay; this would create an endogenous upper bound on the demand for transaction processing.
%Conflux has a theoretical transaction throughput of $3,200$ \textcolor{red}{3,500} per second~\cite{li2018scaling}. 
%As a day has $86,400$ seconds, therefore, on day $d$ there will be $u(d)\times 3,200 \textcolor{red}{3,500}\times 86,400$ as many transactions. 
Under this estimated model, it will take 718 days until Conflux reaches a network capacity of 50\% and 793 days to reach capacity of~70\%. 

%Undoubtedly, the uptake of Conflux may vary from the model described above in terms of the time needed to reach a specific adoption rate. Ecosystem investments and dApp development may enable a faster uptake. Conflux's smart contract platform is compatible  with Solidity, one of the main programming languages for smart contracts on Ethereum. This compatibility implies that many blockchain dApp developers from the Ethereum community face a shallow  adoption curve for Conflux. 
%In contract, when Ethereum was launched, there was a significantly smaller  community of developers  proficient in Solidity. 
%Hence, it is reasonable to expect a faster user uptake of Conflux comparing to Ethereum. 

\begin{figure}[t]
\begin{center}
\includegraphics[width=7cm]{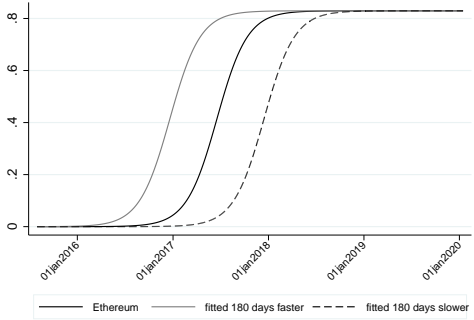}
\caption{The Three Calibrated Adoption Rates\label{eth_uptake_fig2}}
\end{center}
\end{figure}

In calibrating our model, we provide an analysis under two different adjustments to the estimated model as the uptake of Conflux may vary from the model described above in terms of the time needed to reach a specific adoption rate.
First, we shift the adoption curve 180 days to the right, meaning that adoption is delayed by a quarter.
Second, we shift the adoption curve 180 days to the left, meaning that adoption is sped up by a quarter. 
Formally, this shift is an increase/decrease in parameter $x_0$ to 870 and 510 calendar days, respectively:
\begin{eqnarray}
      u^{\text{fast}}(d)&=&\frac{0.83}{1+ e^{-0.017\cdot (d-510)}},\label{uptake_fast}\\
    u^{\text{slow}}(d)&=&\frac{0.83}{1+ e^{-0.017\cdot (d-870)}}.\label{uptake_slow}
  \end{eqnarray}
Figure~\ref{eth_uptake_fig2} illustrates the three adoption rate models, labelled as \textit{fast} ($u^{\text{fast}}(d)$), \textit{Ethereum} ($u^{\text{ETH}}(d)$),  and  \textit{slow} ($u^{\text{slow}}(d)$). 
%\TODO{Yuxi(symbol): Sorry for re-commenting this: but want to make sure it's \textit{Ethereum}($u^{\text{ETH}}(d)$), and \textit{slow} ($u^{\text{slow}}(d)$)}

With an uptake rate of $u(d)$, the average daily number of transactions is as follows:
\[T(d)= u(d)\cdot 3,200 D.\]

%todo: may want to move this paragraph around
At capacity, Conflux has a throughput of $3,200$ tps. 
With  a long-run adoption rate of  $u(d)=80\%$,  this amounts to an expected  $2,560$ tps utilization. 
One can also argue that the adoption rate in Conflux may exceed the above estimates. 
%The implicit assumption is that Conflux will attract a wider user base than Ethereum because at capacity it processes recognizably more transactions than Ethereum does today.
%We justify this assumption as following:
Ethereum is arguably at capacity most of the time (see Figure~\ref{eth_uptake_fig} and its  mem-pool of unsettled transactions is non-empty). 
Since Ethereum is   at capacity, there is a limited incentive for developers to introduce new DApps, especially  for enterprise-scale use-cases. 
Conflux's higher throughput mitigates the concerns that the transactions do not get confirmed timely, and since it is compatible with Solidity, developers face a flat learning curve.
Together this should contribute to a fast adoption of Conflux.

% transaction fees subsection
To simplify, we denominate the capacity by the number of native token tps. We assume that users on average pay a transaction fee of value $f$. 
Therefore, average daily fees, as a function of day $d$, $F(d)$, are as follows:
\begin{equation}\label{userfeerevenue}
    F(d):=f \times T(d) = f\cdot u(d)\cdot  3,200 \cdot D.
    \end{equation}
    
\iffalse
  The table below presents examples for the annual fee income on the Conflux Network that ``at capacity'' usage rate provides to miners for different levels of the average fee  $f$. 
 
 \begin{table}[ht]
     \centering
 %\begin{center}
     \begin{tabular}{lrr}
\hline\\
transaction fee & transaction fee per day & annual Tx fees \\\\\hline\\
 \$0.001  &  \$221,184\textcolor{red}{241,920}  &  \$80,732,160\textcolor{red}{88,300,800}  \\
 \$0.005  &  \$1,105,920\textcolor{red}{1,209,600}  &  \$403,660,800\textcolor{red}{441,504,000}  \\
 \$0.010  &  \$2,211,840\textcolor{red}{2,419,200}  &  \$807,321,600\textcolor{red}{883,008,000}\\
 \$0.020  &  \$4,423,680\textcolor{red}{4,838,400}  &  \$1,614,643,200\textcolor{red}{1,766,016,000}  \\
 \$0.050  &  \$11,059,200\textcolor{red}{12,096,000}  &  \$4,036,608,000\textcolor{red}{4,415,040,000}  
  \\\\\hline
\end{tabular}
%\end{center}
\end{table}
\fi

For Ethereum, at its current block reward and hash rate, total rewards are
on the order of \$2.3M daily or \$840M annually, including both block rewards and transaction fees~\cite{ethPrice}. 
As a result, transaction fees account for less than 3\% of the rewards.
In Conflux, with a similar block-usage rate, transaction fees would provide the same total fee income as the \textit{total} revenue (fees plus block rewards) in Ethereum as long as user are willing to pay on average \$0.01 per transaction, a desirable feat. 
Even for a moderate willingness of users to pay fees, annual income can be substantial. 
In  comparison,  the  median transaction fee on the Ethereum blockchain for January--February 2020  has been between \$0.08 and~\$0.15~\cite{ethFee}.

\subsection{Storage Bonds and Interest payment}\label{bondedstorage}

\begin{figure}[t]
\begin{center}
\includegraphics[width=7cm]{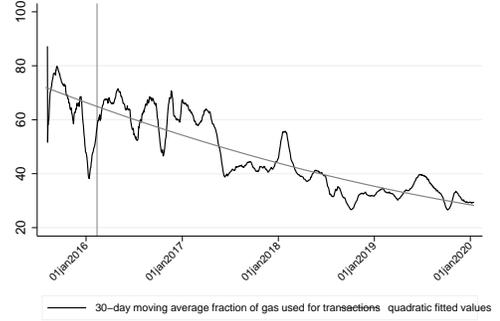}
\caption{Transactions vs.\ Computations\label{trans_fig}}
\end{center}
\end{figure}

To characterize the size of storage bonds, we start with the modeling of transaction fees split by token ownership transfers vs.\ computation. 
Figure \ref{trans_fig} shows the fraction of gas attributable to address-to-address transfers in percentiles in Ethereum.\footnote{We derive this  line as follows. We obtain from \cite{etherscan} the data series for daily transactions and daily Gas used. A simple transfer of ETH transaction requires 21,000 Gas, and we therefore obtain the computation-driven Gas amount by subtracting the number of transactions times 21,000 from the total gas.}
As the figure shows over time simple ownership transfers account for a  decreasing proportion of transactions. 

%\TODO{This figure need to be fixed, legend is unreadable and lacks y axis label}

We characterize the computation-rate with an OLS regression for a quadratic fit in the following form:
\begin{equation}\label{qfit}
    \textit{\% computation gas}=\alpha+\beta_1 \cdot d+\beta_2\cdot d^2+\epsilon,\end{equation}
    where $d$ are the number of days since main-net launch. The goal here is to measure the \textit{\% computation gas} as a quantity $\in[0,100]$. \
%Table \ref{nontx} contains the results for non-linear least squared regression of (\ref{qfit}) using data for Ethereum's the fraction of daily non-transaction gas usage on Ethereum. MA refers to a 30-day moving average. 
%todo: move this to a footnote

\iffalse
\begin{table}[t]
\caption{Quadratic Curve-Fitting for Ethereum's Non-Transaction Gas Usage Rate\label{nontx}}
    \centering
    \begin{tabular}{lll}
\hline\\
& \% non-transactions & MA(\% non-transactions) \\
\\\hline\\
$\beta_1$ & -0.04*** & -0.04*** \\
 & (-17.952) & (-26.239) \\
$\beta_2$ & 0.00*** & 0.00*** \\
 & (5.598) & (7.710) \\
$\alpha$ & 71.88*** & 72.07*** \\
 & (94.875) & (141.995) \\
 \\
Observations & 1,623 & 1,623 \\
R-squared & 0.615 & 0.782 \\\\\hline

\end{tabular}
\end{table}
\fi

Specifically, let $\gamma(d)$ denote the fraction of gas usage for computation. Following the quadratic fit, we obtain:
\begin{align}\label{nontx_eq}
    \gamma(d)&:=1-\left(72-0.04\cdot d+7.05\cdot10^{-6}\cdot d^2\right)/100 \nonumber\\
    &=-0.0000000007(d-2,837)^2+.85. 
\end{align}
We note that the parameter estimated for the quadratic term, $\beta_2$, is very small, around $7.05\times 10^{-6}$, owing to the size of the associated variable.
Therefore, when there are $T(d)$ transactions on day $d$, we say that $(1-\gamma(d))\cdot T(d)$ of these are token ownership transfers and $\gamma(d)\cdot T(d)$  involve smart contract executions that require data storage on the chain.
%So far in this subsection, we have discussed the usage of gas for computation purposes. In what follows, we need to determine the amount of interest that miners receive for storing smart contract code. We measure miner income against transactions because transaction fees are paid per transaction and a block has a fixed capacity in terms of number of transactions. A more intuitive approach would be to measure this quantity against gas usage. To keep the analysis coherent, we convert the fraction of computation gas into hypothetical transactions and then we  make the interest payments a function of these hypothetical transactions.

%figure 5: to force it on top of page 6
\begin{figure*}[t!]
\begin{center}
\begin{tabular}{ccc}
\includegraphics[width=5.5cm]{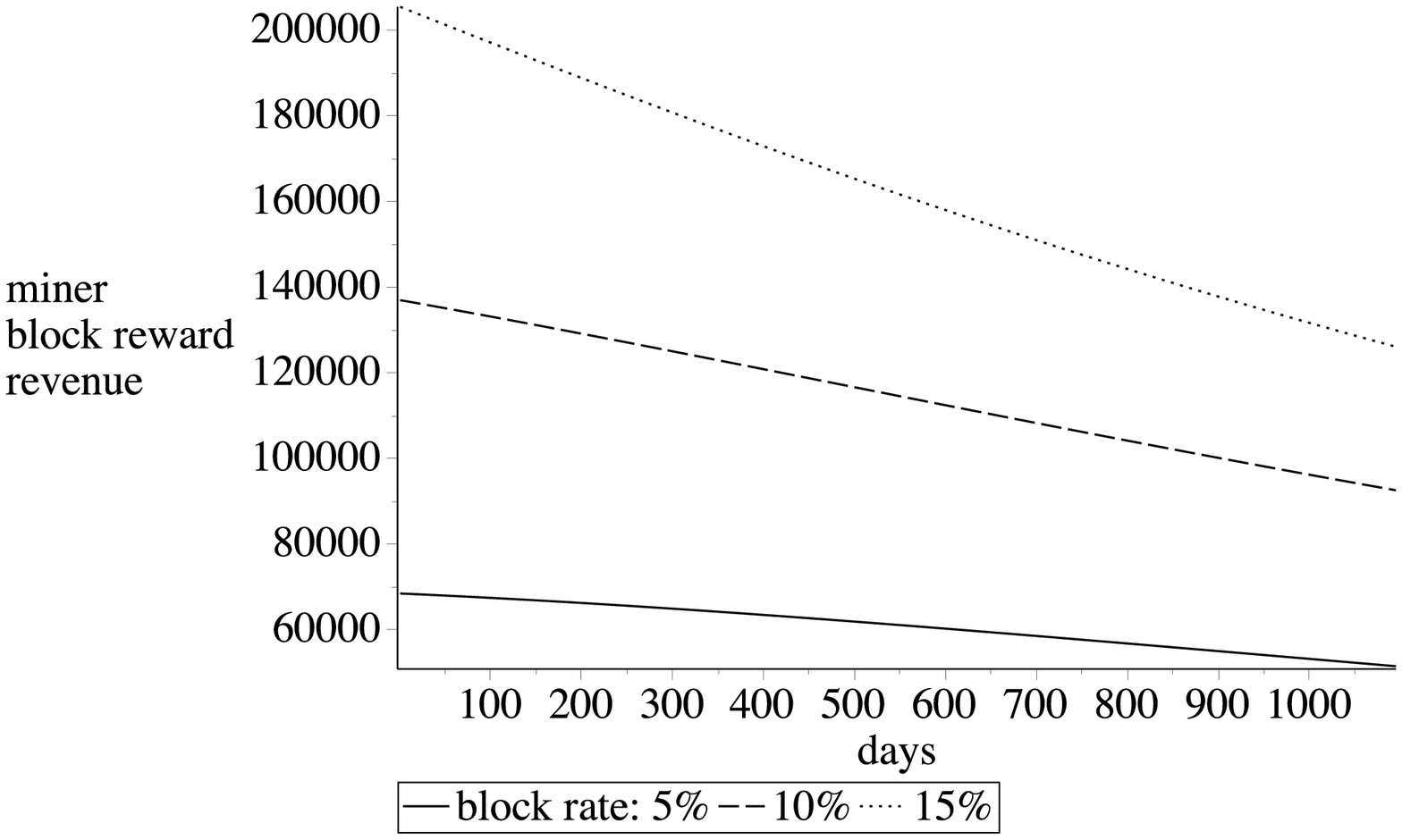}
& \includegraphics[width=5.5cm]{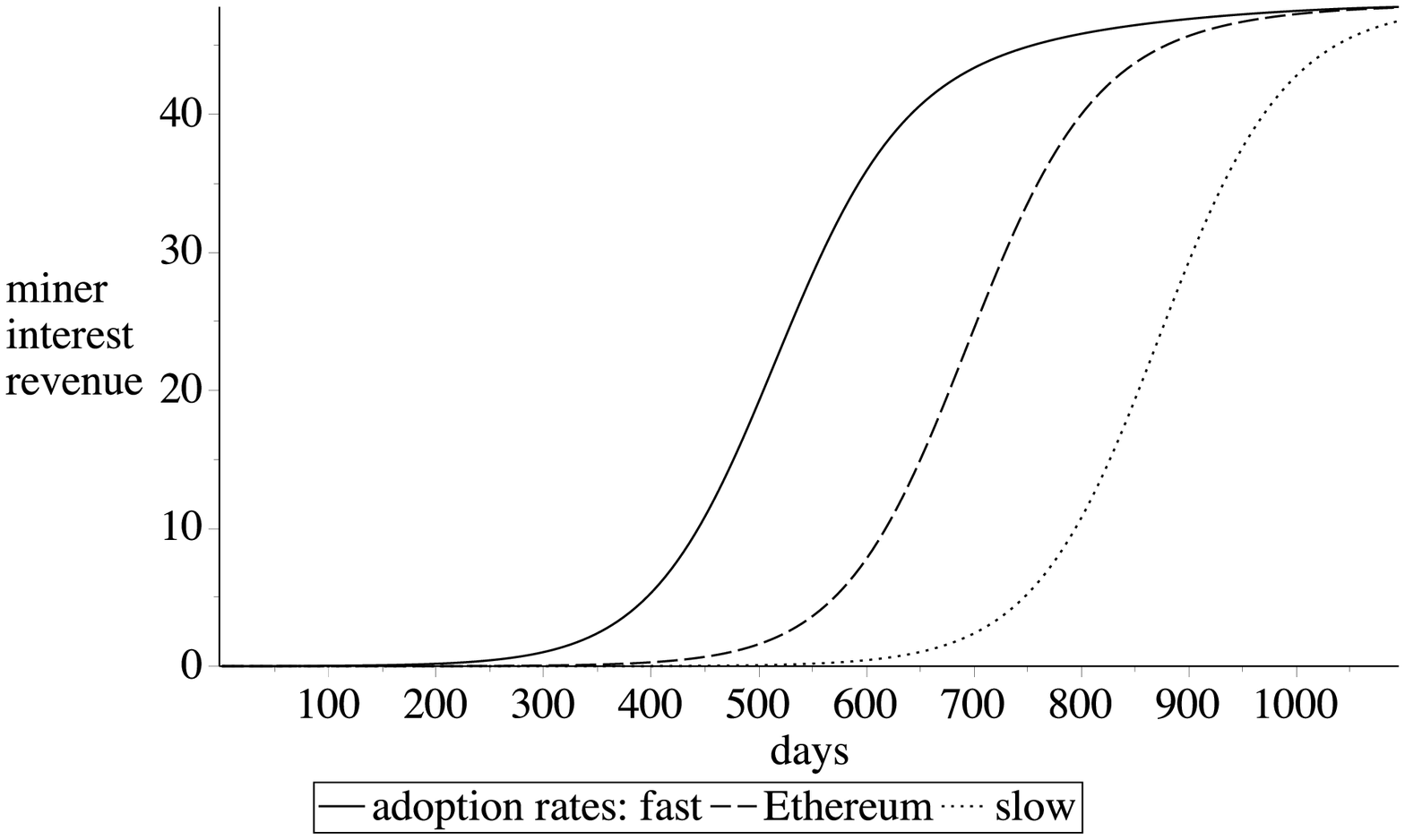}
& \includegraphics[width=5.5cm]{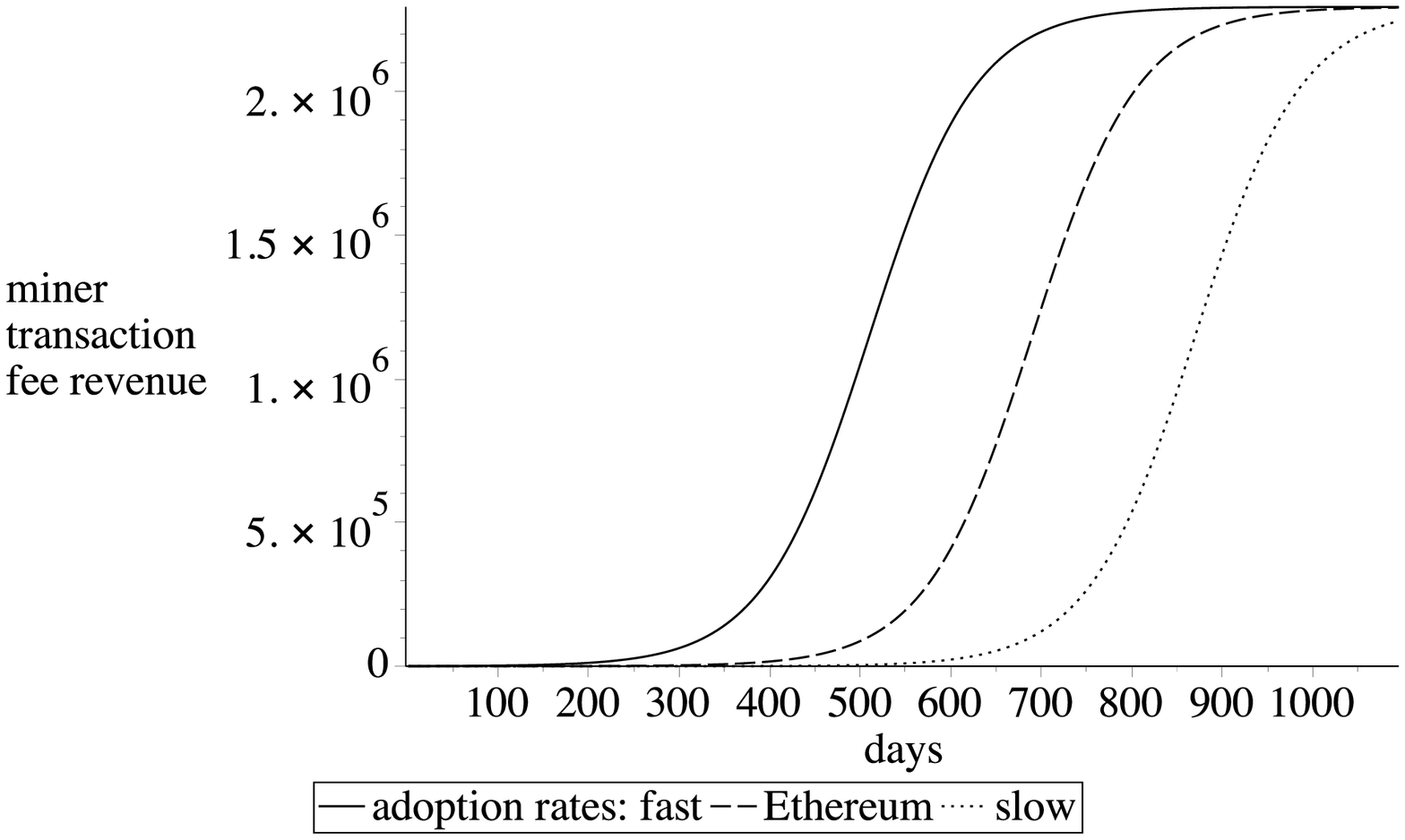}\\
\textit{Panel A: Block Rewards}
&\textit{Panel B: Interest Income}
&\textit{Panel C: Transaction Fee Revenue}
\end{tabular}
\end{center}
\caption{Miner Revenues over time as a Function of the Adoption Rate\label{uptake_revenue_fig2}}
\end{figure*}

%\subsection{Interest Payments to Miners}
%This section estimates the interest payments to miners in Conflux. 
To simplify the interest payment estimation, %In calibrating the model, there is a possible scenario that a user buys storage ({\em i.e.,} put tokens as storage bonds) but never executes the contract hereafter. 
we make the assumption that users make the decision of whether to stake tokens as storage bonds each day and, therefore, that the total transactions fully reflect the extent of interest payments. This rules out a possible scenario that a user buys storage ({\em i.e.,} put tokens as storage bonds) but never executes the contract hereafter. 
In other words, we account for only ``new'' bonding of tokens. 
As such, the calibration model likely {\em conservatively underestimates} the interest income to miners.

The required storage bonds are proportional to the size of the contract code.
We assume that this amount is proportional to the gas usage of the contract or, as one may argue,  the number of  actual transactions since each of them requires gas. 

For $x$ transactions, users need to put $\beta\cdot x$ tokens as storage bonds and on day $d$ it is $\gamma(d)\cdot T(d)$ transactions that require storage bonds. 
In total, the required amount is $\beta\cdot\gamma(d)\cdot T(d)$. 
We conclude that each day the miners receiving  interest  paid on these storage bonds is:
\begin{eqnarray}\label{interest_miner}
I(d):=\beta\cdot \gamma(d)\cdot T(d) \cdot R,\end{eqnarray}
where $R$ represents the daily interest rate for compound transactions. 
%Details are in~\cite{YCAPAV:2020}. 
%\TODO{You cannot reference something which is not even archived, and it's an internal document, so you need make the PAR here self-sufficient and delete claims you do not back up with math}

\subsection{Total Miner Revenue}\label{totalminerrevenue}

To summarize, total miner revenue, denoted by $m(d)$, consists of {\em (a)}~the  block reward from equation~(\ref{blockrewards}), {\em (b)}~transaction fees expressed by equation~(\ref{userfeerevenue}), and {\em (c)}~interest income from bonded tokens as shown by equation~(\ref{interest_miner}):
\begin{eqnarray}\label{total_miner_revenue}
    m(d)&=& p(d)\cdot b(d)+ F(d) + p(d)\cdot I(d).
\end{eqnarray}
Absent exogenous forces that affect the  market price, the price of CFX token on day~$d$, $p(d)$, is determined simply by the total number of tokens outstanding, $p(d)=$initial price $\times$ genesis tokens$/($genesis tokens $+$ block rewards $+$ interest payments$)$. 
%The newly minted block rewards and interest payment tokens can be determined in closed form; see~\cite{YCAPAV:2020}.
%\TODO{See commment above. Claims need be self sufficient so delete things we do not back}

%\subsection{Calibration of Miner Revenue}

%In calibrating the numbers we aim to generate an average daily miner revenue that mimics what miners can currently earn on Ethereum, which is about \$12.50 per giga-hash.
%Before presenting the calibration results for Conflux miner revenue, we determine the  Ethereum miners' revenues as a benchmark. 

Before we present our calibration results for mining revenue on Conflux, as a benchmark we set how much Ethereum miners earn.
There are around 6,500 blocks created per day, paying around 13,500 ETH so that the total average daily block rewards is around \$3M USD (at current ETH/USD prices)~\cite{ethPrice}.

\iffalse
For Ethereum, transaction fees  are negligible for  miner income, whereas for Conflux transaction fees are expected to play a more significant role due to the high tps throughput.
Since the network first needs to gain traction, there will be few transactions early on. 
Therefore, when presenting our  calibration results we average miner income over a longer horizon, namely,  we compute year-long averages in 91-day intervals:
 \[\bar{m}(d)=\frac{1}{365}\sum_{d'=i\cdot 91+d}^{(i+4)\cdot91+d} m(d'),~~~i=0,\ldots,7.\]
\fi

We use four values for average transaction fees, $f\in\{.005,.01,.02,.08\}$, where the highest number \$.08 corresponds to the low-end median fee paid on Ethereum in early 2020, as we discussed earlier.
For the uptake rate, we consider the three benchmark rates $u^{\text{fast}}(d)$, $u^{\text{ETH}}(d)$, and $u^{\text{slow}}(d)$ from Subsection~\ref{uptake}.
For the storage bond requirement, we use $\beta=1\%$ meaning that if the user occupies space on the blockchain for future computation that is equivalent to what one  virtual-machine opcode transaction occupies, then this user has to put 1/100 of a CFX token into bonded storage. 
We also assume that there are no exogenous market-driven price changes except where explicitly stated.

% figure 6
\begin{figure}[t]
\begin{center}
\includegraphics[width=7cm]{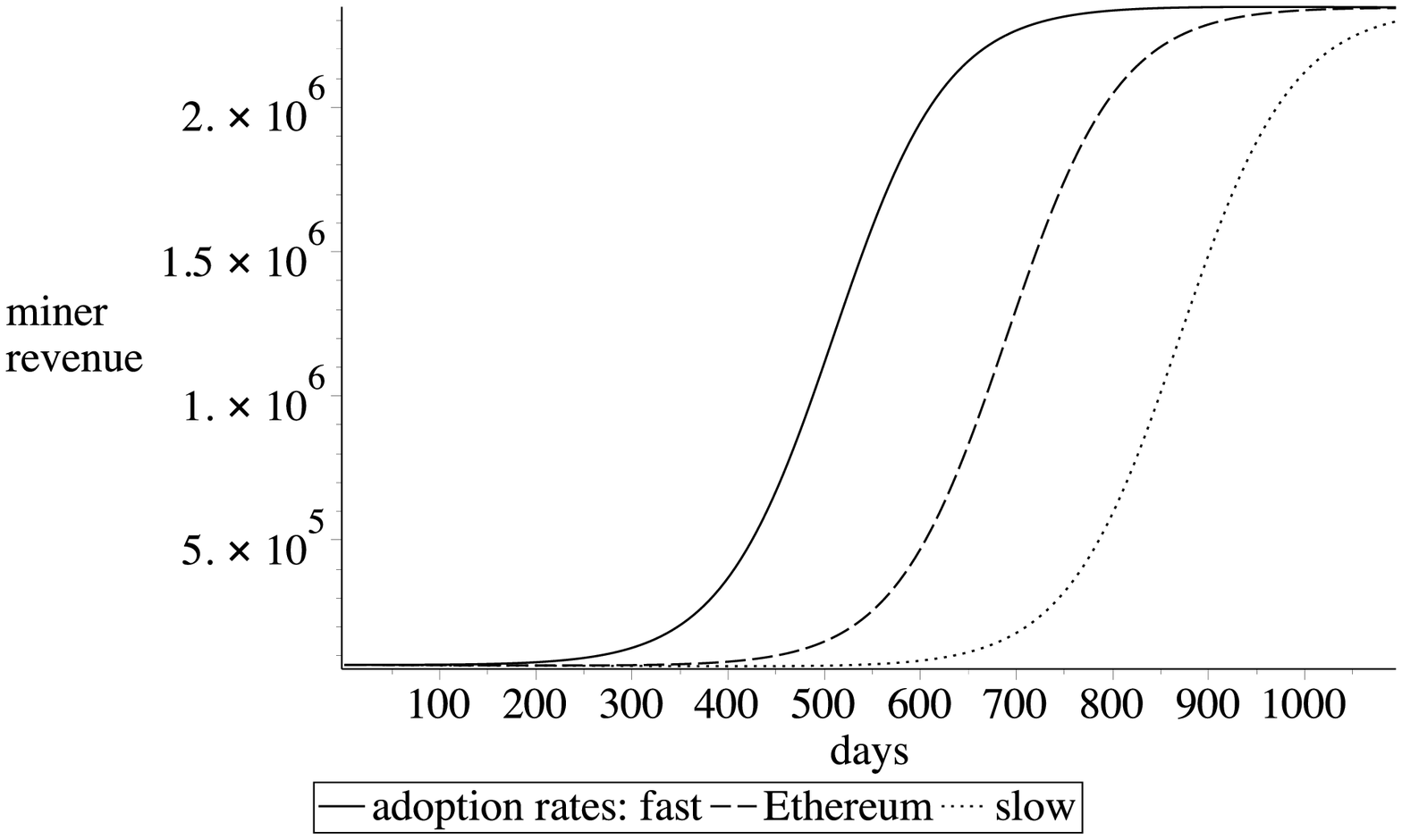}
\end{center}
\caption{Miner Revenues over time as a Function of the Adoption Rate\label{uptake_revenue_fig}}
\end{figure}

Figure \ref{uptake_revenue_fig2} plots the three components of miner revenue: block rewards, interest income, and transaction fees.
These figures use an annual interest payments of $r_c=2\%$, and average fees of \$.01.
The \$-value of block rewards (Panel A) declines because the price declines due to inflation; note that we assume that the number of tokens given as a block reward is constant within the interval. 
For the remaining two panels, we set the annual block inflation rate to $r_b=5\%$.
Interest income (Panel B) rises with blockchain usage, but it is small in magnitude. 
Finally, transaction fee revenue (Panel C) plots fee income.
The values recorded on the vertical axis indicates that these fees are expected to be 
an order of magnitude larger than interest income or block reward income, except immediately after the launch of the main-net.

Combining these three figures, Figure \ref{uptake_revenue_fig} plots expected daily miner revenue $m(d)$ over three years following the launch of the main-net for the three different user uptake speed scenarios.
This figure uses an annual block inflation rate of $r_b=5\%$, annual interest payments of $r_c=2\%$, and average fees of \$.01.

\begin{figure}[t]
\begin{center}
\includegraphics[width=7cm]{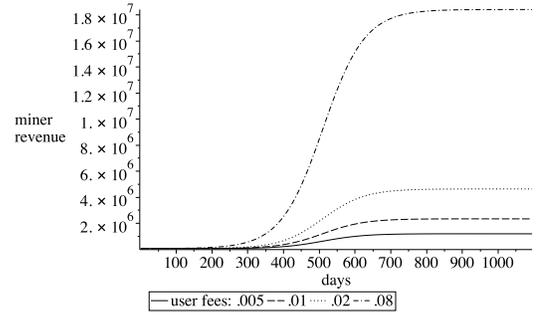}
\end{center}
\caption{Miner Revenues over time as a Function of Average Fees\label{fee_revenue_fig}}
\end{figure}

Figure \ref{fee_revenue_fig} shows the time series of expected miner revenues per day with the four different average transaction fees.
When Conflux is at capacity, even for moderate fees of \$.02, miner revenue will be around~\$4.6M.
This figure uses block inflation rate of 5\%, interest payments of 2\%, and Ethereum-like adoption rates. 

For the sake of the argument, we also consider a situation when market forces lead to increases in the price of CFX tokens such that in three years Conflux has the same market valuation as Ethereum today, that is, roughly a \$15B market-cap.
Further assume that the price change follows linear growth at some 
rate $g$ such that the price at time $d$ is $p^\text{ETH}(d)=p(0)\cdot (1+g)^d$.
The rate  $g$  that ensures that the market evaluation of Conflux three years after launch is the same as Ethereum at the beginning of 2020 is $g\approx0.0031$. 
We compute total miner revenue for the ``speculative'' price $p^\text{ETH}(d)$, i.e., in~(\ref{total_miner_revenue}), we substitute $p(d)$ with $p^\text{ETH}(d)$ so that:
\begin{eqnarray}\label{total_miner_revenue_ETH}
    m^\text{ETH}(d)&=& p^\text{ETH}(d)\cdot b(d)+ p^\text{ETH}(d)\cdot I(d)+F(d)~
\end{eqnarray}

\begin{figure}[t]
\begin{center}
\includegraphics[width=7cm]{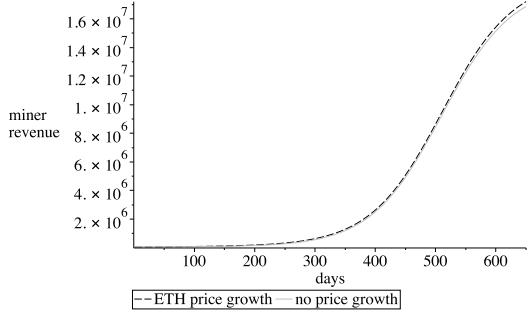}
\end{center}
\caption{Miner Revenues if prices would grow to ETH levels\label{ETH_revenue_fig}}
\end{figure}

Figure \ref{ETH_revenue_fig} shows  the time series of expected miner revenues per day for this alternative price path, $p^\text{ETH}$, where we plot only the  first 650 days.
In this Figure, we use a block inflation rate of 5\%, interest payments of 2\%,  Ethereum level adoption rates, and willingness to pay fees at current Ethereum rates (\$0.08). 
We also include the revenue case when there is no price growth (it corresponds to the most ``optimistic'' case in  Figure~\ref{fee_revenue_fig}) as a point of reference.
%\TODO{Yuxi(presentation): not sure what is compared.}
The {\em key} takeaway from this figure is that when we assume that prices rise significantly, miner income in the medium run is not affected, simply because transaction fees continue dominate.

We conclude that early on, block rewards play the most important role in miner income at the beginning, whereas, once a certain adoption rate is reached, transaction fees will be the most important source of income. 
We emphasize, however, that this is not to say that interest is irrelevant for user decisions.
Instead, there will be many users who each have to pay a  small but possibly for their case significant implicit fee for storing data on the network.

%auto-ignore

\section{Economic Limits against Attacks}\label{sec:limits}
In this section, we examine the limits of the Conflux network under two different attacks, the selfish mining attack and the double-spending attack. 

\subsection{Selfish Mining Attacks} 
If a participant in Bitcoin holds more than 23.21\% of the network computation power, she can
gain more mining profit by strategically withholding her mined block for a period of time before broadcasting them to the network~\cite{sapirshtein2015optimal}. This is
because Bitcoin only gives reward to the blocks in the longest chain. When she withholds the newly mined block, she has the exclusive privilege to mine under her new block which is the current longest chain. Of course, withholding the block brings the risk that someone else may mine a new block concurrently to become the new longest chain, but the study shows that if the participant has more than 23.21\% of the network computation power, the benefit of withholding will outweighs the risk~\cite{sapirshtein2015optimal}. Because Bitcoin mining is a winner-take-all game, honest miners expect to get less reward comparing to their computation power when the selfish participant launches such fairness attacks.

\begin{figure}[t]
\begin{center}
\includegraphics[width=8cm]{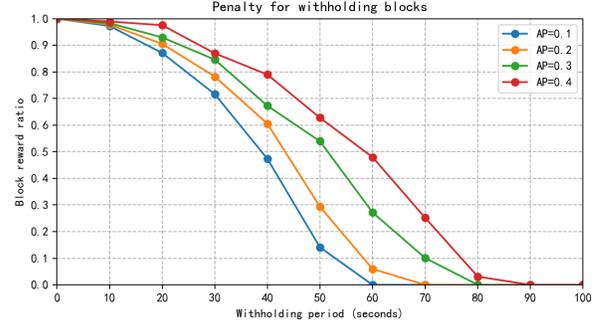}
\end{center}
\caption{Penalty of attackers on different attacker ratios of block generation power (AP)\label{attackPenalty}}
\end{figure}

Conflux is more resilient against selfish mining attacks because withholding a block leads to less reward. Unlike Bitcoin, all blocks receive a reward in Conflux and the reward of a block is discounted by its anti-cone size. Withholding the block will prevent future blocks from referencing it. Therefore, it increases the anti-cone size of the block and consequently decreases the block reward. Given all network participants are rational, honest mining is incentive compatible.

Figure~\ref{attackPenalty} presents our experimental results to illustrate the resilience of Conflux against selfish mining attacks. We run a Conflux network simulation with 10000 nodes. One of them is the attacker which will withhold her generated block for a certain period of time. In the simulation, normal nodes have the network delay (4.1 seconds in average). 
The attacker, however, has the capability of instantly receive and send its block to all other nodes. We run the simulation for 2000 blocks and measure the reward ratio the attacker receives comparing to the normal honest strategy for the last 1000 blocks under different the block generation power and the block withholding period.
Our results show that the attacker consistently receives less reward than she would with the normal honest strategy (i.e., the reward ratio is less than 1). The longer she withholds the blocks, the less reward she will receive. More computation power will help the attacker to receive more reward, but even with 40\% of the computation power of the whole network, the attacker would still get more reward if she just participates the network honestly.

\if 0
Firstly, we consider the scenario that the attacker controls less than $50\%$ of the total computation power. On a serial blockchain, the chance of a successful double-spending attack under such a setting decreases exponentially in the number of block the attacker is behind~\cite{nakamoto2012bitcoin,Sompolinsky:2015}. In Conflux, recall that reward for a block is discounted by the anti-cone size of the block, therefore selfish mining is punished through a decremented block reward. 
We denote the marginal punishment on concurrent blocks as $\gamma$ (\textit{i.e.}, the penalty of a block in the anti-cone). Evidently, block reward of a withheld block is strictly less than a block broadcasted on its generation. 
Figure \ref{attackPenalty} shows that withholding a block results in strictly less block rewards comparing to announcing the block honestly with $\gamma = 0.005$. The penalty increases with the withholding period and is heavier for attackers controlling less power. %Figure \ref{} illustrates how such a penalty affects the comparative fairness. The vertical axis indicates ratio of the actual reward share of the attacker comparing to the honest miners. The result indicates that when $\gamma$ is large, even withholding is not applied, attackers can take comparatively more rewards than honest nodes because their network advantage relative to honest nodes makes them suffer less from the penalty on concurrent blocks.
\fi

\subsection{Double Spending Attacks}
Several works in the economics literature highlight that PoW networks face fundamental constraints in terms of the economic incentives that can sustain ongoing security of the network~\cite{BIS:2019}.
The Conflux network is no different but in what follows, we argue that the constraints of Conflux are  ``looser'' when compared to existing networks. 
In this section, we make the reasonable assumption  that an attacker is not capable of reversing cryptographic functions, therefore honest miners behave correctly even with the presence of an attacker. 
We focus on double-spending attacks with selfish mining through withholding of blocks.  

We first repeat the arguments from  \cite{budish:2018} which apply to serial blockchains.
We assume that the mining of each block involves a cost $c$ (including physical equipment and electricity) and that there are $N$ identical miners who compete.
For the scenario with negligible user fees, the most significant revenue is the block reward $B$ per block. 
The miners' participation constraint  requires the expected gain to exceed the expected~cost, that is:
\[\text{probability of winning the block}\times B\ge \text{cost}~~\Leftrightarrow~~B/N\ge c.\]
This condition holds for all identical miners, and in equilibrium it must hold that the aggregate cost of mining agrees with the aggregate benefit:
\begin{equation}\label{IC}
c\times N=B.
\end{equation}
Now suppose an attacker wants to double-spend a transaction of value $V$.
The attack proceeds in the sense that the attacker builds an alternative chain faster than all remaining miners.
Assume that to gain $50\%$ power, the attacker has to pay $c\times N$, and to gain a majority they have to pay in excess of this.
If the attacker spends $A\times c\times N $ on equipment, with $A>1$, they gain an advantage of $A/(A+1)>50\%$; the larger $A$, the  larger the advantage (and thus the faster they finish the attack).  
For a successful attack, they earn value $V$, which is the amount that they can double spend.
Assume that, conditional on the equipment advantage $A$, it takes $t$ blocks (in expectation) to complete the attack, that is creating a longer chain than the chain honest miners collaboratively generating.
Then the cost of the attack is:
\[t\times A\times c\times N.\]
Once successful, however, the attacker earns not only the attack value $V$ but also rewards for the $t$ blocks.
Therefore, for attacks to be {\em unattractive}, it must hold that:
\begin{equation}
t\times A\times c\times N>V+t\times B.
\end{equation}
Using equation (\ref{IC}), we obtain the following:
\begin{equation}\label{ec1}
t\times B(A-1)>V.
\end{equation}
Therefore, for an expected attack time $t$, there exists a value $\cal{V}$ such that for all $V>t\times B(A-1)=\cal{V}$, and the transaction of value $V$ cannot be secured.
Inequality (\ref{ec1}) is a firm constraint  on the economics (and the security) of a serial chain such as Bitcoin.

Conflux subjects to a different lower bound for $V$.
First, to be successful in an attack, the attacker's alternative chain must become the pivot chain.
Since any epoch may contain multiple blocks, not only the attacker needs to create blocks faster, but also to generate a ``heavy'' chain, which will require relatively more time (and thus more resources).
To simplify the argument, we  abstract from this issue and assume, as before, that the honest chain contains a single block per epoch.

Next, when creating the alternative chain, an attacker does not receive the full reward because block rewards are assigned based on the block's anti-cone size. 
As before, suppose there is a single attacker in the system, who succeeds an attack in $t$ blocks. 
Assume that the attacker references honest block as soon as one is seen, the attacker's first block in the alternate chain has an anti-cone of size of at least $t-1$, the second of $t-2$, and so forth. 
Therefore, the block reward for block $a$ since the start of the attack is $B\times\left(1-\left(\min\{t-a,10\}/100\right)^2\right)$ assuming a fixed per block reward $B$.
For the longest chain (now the pivot chain) of length $t$ since the start of the attack, the attacker will therefore earn:
\[B\cdot \underbrace{\sum_{i=1}^t \left(1-\left(\frac{\min\{t-i,10\}}{100}\right)^2\right)}_{\Pi_t}<t\times B.\]
Using the same argument as above,
and therefore, the economic constraint for Conflux becomes:
\begin{equation}
B(tA-\Pi_t)>V
\end{equation}
In other words, there exists a value  $\cal{V}'$ such that for all $V\in(\cal{V},\cal{V}']$, 
the following holds: 
\[B(tA-\Pi_t)>V>B(tA-t)\] 
The implication of this relationship is that the set of transaction values $V$ that can be secured on the Conflux network is strictly larger than in ``traditional'' serial blockchains such as  Bitcoin under such an attack strategy.

\vspace*{-4pt}
%auto-ignore

\section{Conclusion} \label{sec:conclusion}
The long term sustainability and economic resilience to attacks are critical to a decentralized, proof-of-work blockchain network. 
When basing our economic
calibration on similar uptake and usage of Conflux as of Ethereum, we observe that as adoption increases, the significantly higher throughput of the network allows user fees and storage interest payment to make up the bulk of income for miners, making the mining activity sustainable in the long term. Our analysis results also show that Conflux with its novel incentive mechanism is more resilient when facing double-spending attacks and selfish mining attacks than sequential blockchains.

\if 0
Miners play a crucial role in the operation of a decentralized, proof-of-work blockchain network. 
The amount of computing power and thus network security is increasing in the revenue that they can expect from their activities.
We assess here which level of income and thus network security the  Conflux network, a new proof-of-work blockchain, can generate, and how it depends on user behavior and ``policy variables'' such as block and interest inflation. 
We also discuss how the engineering design extends  the economic boundaries compared to  other PoW blockchains. 
When basing our calibration on similar uptake and usage of Conflux as of Ethereum, we observe that in the early stages after main-net launch, block  rewards  in Conflux will play  the  most important  role  in  miner  income. 
As adoption  increases, the significantly higher throughput of the network allows user  fees to make up the bulk of income for miners. 
Additionally, Conflux is theoretically more resistant to double-spending attacks than  other sequential blockchains. 
\TODO{Needs summarize better the paper}
\fi

%\section*{Acknowledgment}
%This work was supported by the Natural Science and Engineering Research Council of Canada.

\ifCLASSOPTIONcaptionsoff
  \newpage
\fi


\Urlmuskip=0mu plus 1mu\relax
\bibliographystyle{IEEEtran} 
\vspace*{-4pt}
\bibliography{BRAINS20}

\end{document}